\documentclass[twocolumn]{revtex4-1}

\usepackage{graphics} 
\usepackage{graphicx} 
\usepackage{color} 
\usepackage{amsmath}
\usepackage{amssymb}
\usepackage{amsfonts}
\usepackage{sidecap}

\begin{document} 

\title{Rich polymorphic behavior of Wigner bilayers}

\author{Moritz Antlanger$^{1, 2}$} 
\author{Gerhard Kahl$^1$} 
\author{Martial Mazars$^{2}$} 
\author{Ladislav \v{S}amaj$^{3, 4}$} 
\author{Emmanuel Trizac$^4$} 

\affiliation{$^1$Institute for Theoretical Physics and Center for Computational Materials Science (CMS), Vienna University of Technology, Wien, Austria \\ 
$^2$LPT (UMR 8627), CNRS, Univ. Paris-Sud, Universit\'e Paris-Saclay, 91405 Orsay, France,\\ 
$^3$Institute of Physics, Slovak Academy of Sciences, Bratislava, Slovakia \\ 
$^4$LPTMS, CNRS, Univ. Paris-Sud, Universit\'e Paris-Saclay, 91405 Orsay, France}


\begin{abstract}
Self-assembly into target structures is an efficient material design strategy. Combining analytical calculations and computational techniques of evolutionary and Monte Carlo types, we 
report about a remarkable structural variability of Wigner bilayer ground states, when charges are confined between parallel charged plates. Changing the inter-layer separation, or the 
plate charge asymmetry, a cascade of ordered patterns emerges. At variance with the symmetric case phenomenology, the competition between commensurability features and charge neutralization 
leads to long range attraction, appearance of macroscopic charges, exotic phases, and non conventional phase transitions with distinct critical indices, offering the possibility of a subtle, 
but precise and convenient control over patterns.
\end{abstract}
\date{\today}

\maketitle

The self-assembly of colloidal systems opens the way to the synthesis of materials that considerably widen the class of known natural crystals, among which opals or butterfly wings. From an 
academic perspective, these new, complex structures allow for detailed and original studies of fundamental processes like nucleation, glass transition or low dimensional statistical 
physics \cite{Pert01,DHLM13}. Skillfully combined with progress in particle synthesis, self-assembly has led to a wealth of applications such as patterned magnetic systems or bandgap 
materials used in displays, optical devices, photochemistry and biological  sensors \cite{WhGr2002,MdGBD11,SBSP12,Abecassis14}. Taking advantage of targeted self-organization requires 
a fine control of interactions between the entities under study. This tailoring is achieved in most cases either by a) introducing some patchiness on the colloids, b) increasing the complexity 
of the problem, by considering e.g. mixtures instead of pure systems, c) affecting the solvent through various  additives (polymeric, electrolytic, etc.), d) introducing an external field, 
be it electric, magnetic, laser-optical or stemming from the interactions with a patterned substrate \cite{Bech08,Lowen08,Glotzer07,Granick11,Bianchi_ACS_2013,Bianchi_2014}. As fruitful as  
they have turned out to be, these strategies in general do not allow for convenient {\it in situ}\/ changes of the obtained ordered structures, so that it is challenging to probe and tune their 
variety in a simple fashion, by controlling an external parameter. Relinquishing the four routes above, we consider here a pure classical system of charges and show that the simplest form of 
external control -- confinement in a slab -- induces an unexpected  structural variability, which in turn opens the way for a precise structural control. We shall focus on energy minimizing 
configurations, relevant when the kinetic energy is small compared to the Coulombic potential energy, and where the charges are forced into a bilayer configuration, thereby creating a 
particular realization of a so-called Wigner crystal.

Wigner crystals were first predicted by the eponymous physicist in the 1930s for electrons in a metal \cite{Wigner1934}, where they have actually never been observed. Instead, their occurrence 
has been reported in the 1970s for electrons at helium interfaces \cite{GrAd79}. Found in neutron stars and in the interior of white dwarfs, they have subsequently been evidenced in 
semi-conductors \cite{TsSG82,EiMD04,WZYE07,ZHDK14}, graphene \cite{AbCh09}, quantum dots, trapped ionic plasmas or other dusty plasmas \cite{MoIv09}, and in the colloidal realm \cite{Pert01}. 
While the symmetric setup is now completely understood \cite{GoPe96,WeLJ01,LoNe07,OgML09,Samaj12_2}, very little is known for the asymmetric  one \cite{RaJo08}. How do charges organize into the 
bilayer structures that spontaneously form in our problem? Upon answering this question, we will treat the general asymmetric situation (no mirror symmetry between the two layers), which turns 
out to be considerably more complex than the symmetric case.

\begin{figure*}[htbp]
\begin{center}
\includegraphics[width=18cm,clip=true]{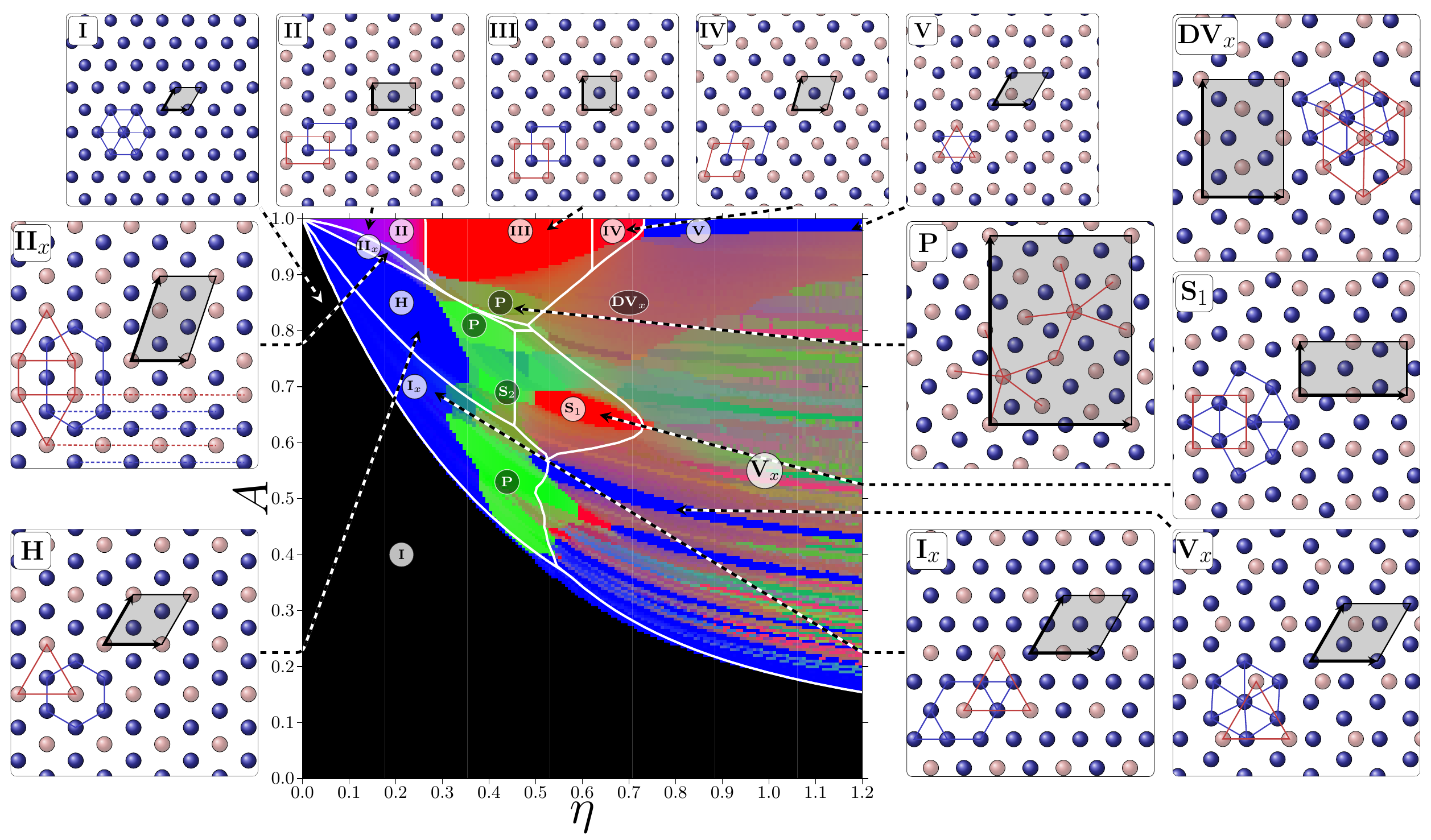}
\end{center}
\caption{Phase diagram in the $(\eta, A)$-plane; $\eta$ measures slab width and $A=\sigma_2/\sigma_1$ is the bilayer charge asymmetry. The color code is that of the so-called RGB scheme 
(red-green-blue); here, the values of 4-fold, 5-fold and 6-fold bond orientational order parameters are added up to yield the color of a given state point: $\Psi_4^{(4)}$ for red, $\Psi_5^{(4)}$ 
for green and $\Psi_6^{(4)}$ for blue. The range of stability of phase I is shown by the black region on the left hand side. It terminates on the rhs at the $A_c(\eta)$ curve. The continuous lines  
are for the analytical predictions of phase boundaries, restricting candidate structures to be of I, I$_x$ (including H), II, II$_x$, III, IV, V, V$_x$, and S$_1$ types. Some typical snapshots of 
structures are shown: those found in the symmetric $A=1$ case, together with I$_x$ (for $\eta=0.184$, $A=0.775$, $x=1/4$), H ($\eta=0.198$, $A=0.85$, $x=1/3$), II$_x$ ($\eta=0.148$, $A=0.95$, 
$x=2/5$), snub S$_1$ ($\eta=0.622$, $A=0.675$, $x=1/3$), pentagonal P ($\eta=0.381$, $A=0.85$, $x=3/7\simeq 0.429$), V$_x$ ($\eta=0.707$, $A=0.5$, $x=1/4$)
and DV$_x$ ($\eta=0.7$, $A = 0.75$, $x=2/5$). For all structures, the unit 
cell is the shaded region. Particles colored red are in plate 2, and those in blue are in plate 1.}
\label{fig:phd}
\end{figure*}

The question addressed is the following. Consider an ensemble of mobile point  charges (``particles'') interacting via a $1/r$ potential, confined between two parallel plates bearing uniform 
charge densities $\sigma_1 e$ and $\sigma_2 e$, with $-e$ the (elementary) charge of the mobile particles; the system as a whole is electroneutral. What is the energy-minimizing arrangement of 
particles for a fixed plate-to-plate distance $d$?  The Earnshaw theorem \cite{Earnshaw} provides a first clue: given that a classical system of point charges under the action of direct (i.e., 
not image) electrostatic forces alone cannot be in an equilibrium configuration, the particles are expelled from the slab interior, and have to stick to the confining plates. Numerical 
and analytical work have furthermore shown that when $\sigma_1=\sigma_2$, staggered configurations arise on each plate, which -- depending on $d$ -- can be rectangular/square, rhombic or 
hexagonal \cite{GoPe96,WeLJ01,LoNe07,OgML09,Samaj12_2}, see also the line $A=\sigma_2/\sigma_1=1$ in Fig. \ref{fig:phd}.
 
Our interest focuses on the asymmetric case ($\sigma_1 \neq \sigma_2$), first on the cornucopia of ordered structures that appear as energy-minimizing, and second, on the distinct properties 
that characterize these new phases. Indeed, macroscopic charges do emerge, which result in a long range attraction between the plates. In addition, different universality classes are probed by 
changing solely the interplate separation, and overcharging is reported in some pocket of the phase diagram. By a combination of complementary analytical and computational techniques of evolutionary 
and Monte Carlo type \cite{suppl}, our goal is to unravel these properties, while charting out the phase diagram.

Without loss of generality, we assume $\sigma_1>0$.  We introduce the asymmetry parameter $A = \sigma_2/\sigma_1$, and consider  $A \in [0, 1]$ \cite{rque23}. We then define the dimensionless 
distance  $\eta = d \sqrt{(\sigma_1 + \sigma_2)/2}.$ Our system is thus entirely specified by $\eta$ and $A$. We further introduce the surface particle densities $n_1$ and $n_2$ and the order 
parameter $x = n_2/(n_1 + n_2)$. Electroneutrality imposes $\sigma_1 + \sigma_2 = n_1 + n_2 $. In general $n_i \ne \sigma_i$ ($i = 1, 2$) and thus each of the plates as a whole (i.e. particles 
plus surface charge density) is charged. $\phi(z) = - 2\pi e (\sigma_1-\sigma_2) z , \, 0<z<d $. If local neutrality holds for both plates, then $n_i$ = $\sigma_i$ ($i = 1, 2$) and we find 
$x=x_{\rm neutr} \equiv A / (1 + A)$. This should be the case when $d\to \infty$, since violating local neutrality would result  in a macroscopic electric field at large $d$, with divergent energy.

Upon changing $\eta$ in the {\it symmetric} case ($A=1$), it is known that a sequence of five phases (denoted I to V) emerges, consisting of two equivalent, ``ideal''  (i.e. undistorted)  
structures on plates 1 and 2, shifted with respect to one another.
For $A=1$, each of the plates is locally neutralized, and $x=x_{\rm neutr}=1/2$. Increasing $\eta$, the hexagonal Wigner monolayer (phase I) is found at $\eta =0$, then a bilayer with rectangular 
arrangements on each plate (structure II), which transforms into a square lattice (structure III). A staggered rhombic  arrangement (phase IV) and a staggered hexagonal lattice (structure V) are 
subsequently observed, see Fig. \ref{fig:phd} \cite{Samaj12_2}. All transitions are continuous, except IV $\to$ V, of first order \cite{Samaj12_2}.

The analytical work proceeds with the derivation of new series representations for the Coulombic energies of undistorted structures \cite{suppl}. 
This yields the exact Coulombic energy of the structures considered.
On the other hand, the numerical work is two-pronged: 
a first technique, inspired from evolutionary algorithms (EA), identifies the optimal periodic structures among all those that have less than 40 particles per unit cell \cite{suppl}; 
the second line of attack consists in extensive Monte Carlo simulations \cite{suppl,Antlanger:15} on much lager systems ($\sim$ 4000 particles per unit cell).
To quantify order and identify the complex patterns formed, it is indispensable to introduce suitable probes: besides the population index $x$, we have used the bond orientational order parameters of 
symmetry $n=4,5,6,7,8,10,12,18$ and 24, as defined from a Voronoi construction \cite{suppl}. These parameters \smash{$\Psi_n^{(L)}$} take unit value under perfect $n$-fold ordering, and have been 
computed in four different variants: restricting to particles in layer 1 (a choice referred to with index $L=1$), in layer 2 ($L=2$),  by projecting layer 1 perpendicularly onto layer 2 ($L=3$), 
or finally by studying the neighbors in layer 1 of a given particle in layer 2 (index $L=4$). While all four variants lead to compatible results, it appears that the latter choice, 
\smash{$\Psi_n^{(4)}$}, with $n=4,5$ and 6 is particularly conducive to investigating the phase behavior. These parameters have been used to construct the phase diagram in Fig. \ref{fig:phd}, 
in conjunction with a precise RGB scheme \cite{rque10}. Fig. \ref{fig:phd} gathers the results for about 35000 state points obtained with EA computations.
The computational cost for MC simulations is about 200 times higher than for EA, due to the complex treatment of long ranged interactions 
in quasi-2D systems (see Ref. \cite{Mazars:11} and references therein). Consequently, a smaller number of state points can be explored with MC simulation and a careful selection 
of those has to be operated to optimize resources \cite{suppl}. MC simulations show that the structures 
obtained following the EA route are stable; more generally a complete agreement EA-MC is reported \cite{comment_2017}.

\begin{figure}
 \begin{center}
\includegraphics[width=8cm,clip=true]{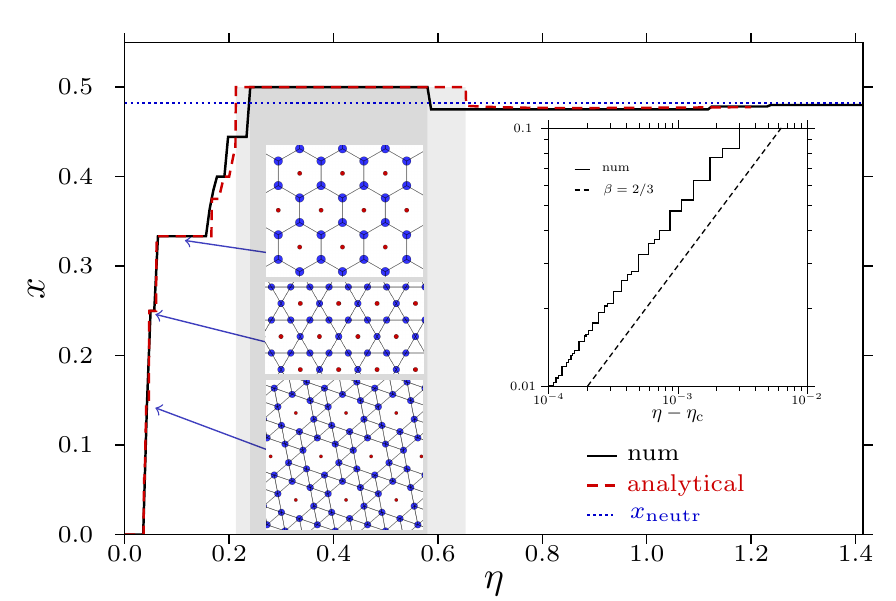}
 \end{center}
 \caption{Order parameter $x$ as a function of distance $\eta$ for $A=0.93$. The horizontal dashed line shows $x=x_{\rm neutr}$ (0.481 here) above which the system is overcharged. The three configurations 
 illustrate the cascade of I$_x$ structures found {(with $x=1/7$, $x=1/4$ and $x=1/3$; from bottom to top)}. As in Fig. \ref{fig:phd}, particles colored red are in plate 2, and those in blue are in plate 1. 
 While the projected pattern (red+blue) is throughout of simple hexagonal type, the partitioning between plate 1 and plate 2 is complex. The inset zooms into the behavior of $x$ in the vicinity 
 of $\eta_c \simeq 0.036$, compared to the predicted power law with an exponent $\beta=2/3$ (dashed line). The shaded grey areas indicate the regions of overcharging (predicted and observed 
 in simulations).}
 \label{fig:x_of_eta}
 \end{figure}

The first noticeable feature revealed by Fig. \ref{fig:phd} is that at small distance $\eta$, it is always favorable for the charges to stick to the plate of highest surface charge (i.e. plate 1). 
Thus, in the black region of Fig. \ref{fig:phd}, the classic hexagonal Wigner monolayer is realized, with $x=0$ (structure I). For large asymmetry (smaller $A$), the monolayer stability is, expectedly, 
augmented. Regions where the system either remains in phase I or partly populates layer 2, are separated by a curve in the  $(\eta, A)$-plane, denoted by $\eta_c(A)$ [or conversely $A_c(\eta)$] shown 
in Figure \ref{fig:phd}.
For $A \gtrsim 0.4085$, this curve separates phase I from a family of phases that are denoted by I$_x$: the latter originates from the monolayer 
(i.e., a hexagonal lattice $\alpha$ on plate 1 with spacing $a$) by picking a fraction $x$ of particles in a hexagonal arrangement to relocate them on plate 2 
(with thus spacing $b > a$). 
An illustration is provided in Fig.\ref{fig:phd}. For these particular 
structures, a complete analytical analysis can be achieved and simple geometric considerations imply that only a discrete set of $x$-values is compatible with this geometric constraint: 
$x=1/(j^2+jk+k^2)$, with non-negative integers $j$ and $k$ such that $j+k>1$:  $x \in \left \{ 1/3, 1/4, 1/7, 1/9,  \cdots \right \}$.
As $x \to 0$, these values become essentially dense, so that we can consider $x$ in this regime as a quasi-continuous variable.
A sufficient condition for instability of phase I is that it becomes energetically favorable to extract one particle from the monolayer, keeping all others in position. This leads to an upper 
bound for $\eta_c(A)$, shown by the thick curve in Fig. \ref{fig:phd}, quite close to the boundary obtained by the EA algorithm. On this curve, the increase in the potential energy of a 
tagged particle shifted from plate 1 to plate 2 is balanced by a decrease in the particle's interaction energy (correlation term). 
For $A \lesssim 0.4085$, phases competing with phase I originate from a different mechanism. This family of phases, denoted by V$_x$, is made up of two triangular (hexagonal) structures on the 
two plates (lattices $\alpha$ and $\beta$) with some shift, see Fig.\ref{fig:phd} for the special cases $x=1/2$ (leading to structure V), and also $x=1/4$. 
Note that when the rescaled distance $\eta \to \infty$, one expects structure 
V$_x$ with $x=x_{\rm neutr}=A/(1+A)$. Considering $x$ as a continuous variable, one can calculate analytically the location of the transition line $\eta_c(A)$, depicted in Figure \ref{fig:phd}, 
along with its EA counterpart.
We present the essence of the calculation, which sheds light on the critical behavior. For a given  $A$-value, the energy difference between structures I and I$_x$ can be written in a small-$x$ 
expansion as
\begin{equation} \label{xinvolved}
\frac{E_{{\rm I}_x}(x,\eta)-E_{\rm I}(\eta)}{e^2 N \sqrt{\sigma_1+\sigma_2}} = f(\eta) x + \frac{2^{3/2}\pi}{\lambda} \eta^2 x^{5/2} + O(x^{7/2}) 
\end{equation} 
where $f(\eta)$ is a closed expression in $\eta$, which also depends on $A$, and $\lambda \simeq 1$ \cite{suppl}. The order parameter $x$ vanishes at the transition, i.e., at a point where 
$\eta =\eta_c$, fixed by the condition $f(\eta_c)=0$. This relation yields the continuous curve $\eta_c(A)$ in Fig. \ref{fig:phd} and can be viewed as a locus of critical points. Besides, 
expanding $f$ for $\eta>\eta_c$ up to linear order in $\eta$ gives access to the critical index. Together with the extremum condition of \eqref{xinvolved} with respect to $x$, this leads 
to the prediction $x \propto (\eta-\eta_c)^{\beta}$ with $\beta=2/3$ \cite{suppl}. This exponent differs from its Ginzburg-Landau theory counterparts, based on an energy expansion that is 
analytic in the order parameter. Here, the long-range nature of Coulomb interaction breaks analyticity. Numerical results are fully compatible with $\beta =2/3$, not only for $A\simeq 1$ 
where it is admissible to neglect lattice distortions (see Fig. \ref{fig:x_of_eta}), but for all $A$ values, along the full curve $\eta_c(A)$. Transitions I$\to$I$_x$ and I$\to$V$_x$ therefore 
share the same non-standard exponent $\beta=2/3$. On the other hand, the other transitions such as II $\to$ III and III $\to$ IV are standard: there, the analytical treatment is rigorous, and 
leads to mean-field continuous transitions, with  $\beta=1/2$ \cite{suppl}. Thus, fixing $A$, it is remarkable that a sequence of transitions with distinct critical indices takes place when 
increasing $\eta$.

Fig. \ref{fig:phd} provides the stability domain of structures I$_x$. Moving away from the $\eta_c(A)$-curve by increasing $\eta$, a cascade of I$_x$-phases emerges, each of them corresponding to 
an $x$-value specified above and associated to a plateau on the left hand side of Fig. \ref{fig:x_of_eta}. A snapshot of structure I$_{1/4}$ is shown in figure \ref{fig:phd}. Noteworthy is the 
honeycomb lattice (H phase) on plate 1, structure  I$_{1/3}$; the corresponding plateau in Fig. \ref{fig:x_of_eta} for $0.07<\eta<0.17$ is of significant extension.
Comparing the boundaries of the H-phase, evaluated via the numerical and the analytical methods shows an excellent agreement for $0 < \eta \lesssim 0.30$, confirming thereby the absence of 
distortions of the optimal structure within that region. For $\eta > 0.45$, the analytical approach still establishes H as the most stable phase (see Fig. \ref{fig:phd}), while EA and Monte 
Carlo identify novel intricate  snub-square or pentagonal structures, see below.

Phases V$_x$ extend over a significant area in the $(A, \eta)$-plane of  Figure \ref{fig:phd}. The agreement numerical/analytical is fair, with discrepancies arising from the emergence of 
complex, distorted structures. At large distances though, where structures $\alpha$ and $\beta$ are undistorted, some exact statements can be put forward \cite{suppl}: (i) for $A < 1$ the 
plates (plus ions in contact) remain charged at any finite distance; only as $\eta\to\infty$ and/or $A\to 1$, does $x$ approach the electro-neutral value  $x_{\rm neutr}$; (ii) while the 
always attractive inter-plate pressure is short-range (exponential) for the symmetric case, it becomes long-range whenever $A\neq 1$, decaying like $1/\eta^2$ \cite{rque24}. A detailed analysis 
of structures V$_x$  reveals that at intermediate distances, they can accommodate significant distortions, leading to new structures coined DV$_x$ in Fig. \ref{fig:phd} \cite{rque2006}.

We now focus on the vicinity of the symmetric line $A=1$, where the structures that prevail for $A=1$ do exist in some parameter range ($A$ larger than $0.9$). These are the $(x= 1/2)$-phases 
II, III and IV.
As the symmetric structures II to IV are undistorted, both the analytical and the numerical approaches predict regions of stability that are in perfect agreement (see the vertical lines in 
the upper part of Fig. \ref{fig:phd}). While for phases III and IV, no generalizations to $x$-values different from 1/2 have been identified, phases II$_x$ emerge in a small pocket of the $(A, \eta)$-plane. 
The numerical EA approach indicates a continuous II$_x \to$ II transition.
The fact that several undistorted II$_x$ structures can be classified via well-defined sequences of alternating rows of particles in the two layers (see Fig. \ref{fig:phd} where the II$_{x = 2/5}$ 
arrangement is depicted), opens the possibility of an analytical, exact energy calculation \cite{suppl}. The numerical and analytical routes provide fully consistent results for the stability of 
these phases.

The upper part of the phase diagram is the locus of a rather unexpected phenomenon of charge reversal. While for $A = 1$ each plate plus ions in contact is electro-neutral  at all distances ($x=1/2$), 
the majority of identified states is characterized by {\it undercharging}: plate 2 (with density $0<\sigma_2<\sigma_1$) plus the (negative) ions in contact carry a net positive charge. Thus plate 
2 attracts {\it less} mobile charges than required for neutrality, so that $x<x_{\rm neutr}$.
This is somewhat expected but is no longer the case for $A \simeq 1$, where {\it overcharging} takes place: the most weakly charged plate attracts {\it more} ions than necessary for neutrality so 
that $x>x_{\rm neutr}$, see Fig. \ref{fig:x_of_eta} \cite{rque25}.

Finally, we report more exotic phases, starting with the snub type. The regular snub square structure S$_1$ shown in Fig. \ref{fig:phd} has $x = 1/3$: particles in layer 1 form a snub square 
lattice \cite{GrSh87}, while particles in layer 2 arrange in a square lattice (with slight deformations as $\eta$ grows). Since particles in layer 1 have five nearest neighbors, the S$_1$ 
phase can be quantified via the five-fold order parameter $\Psi_5^{(1)}$ together with $\Psi_4^{(2)}$ \cite{suppl}. Interestingly, the undistorted snub square lattice is an Archimedean 
tiling \cite{rque52,GrSh87}. Such a geometry is amenable to an analytical treatment \cite{suppl}. Surprisingly, another snub square structure, denoted S$_2$, can be identified with $x = 1/3$ as well. 
In contrast to the S$_1$ phase, it shows stronger deformations, which decrease $\Psi_5^{(2)}$. Both structures occupy relatively small regions in Fig. \ref{fig:phd}, where pentagonal (P) phases 
are also reported \cite{rque30}.

We have considered a charged bilayer system governed by two parameters only: the charge asymmetry $A$ between the parallel plates, and the slab width $\eta$, more prone to experimental tuning. 
The competition of slit confinement with Coulomb interactions leads to a plethora of ordered bilayers with  phase transitions pertaining to different universality classes. In light of the 
simplicity of the model, the complexity and variability of emerging phases is striking. Patterns emerge as a trade-off between the commensurability of structures, and incomplete charge 
neutralization (the latter effect being quantified by $x-x_{\rm neutr}$). Fig. \ref{fig:phd} summarizes our main findings. Besides possible experimental confirmation in quantum wells \cite{WZYE07}, 
semiconductor bilayers \cite{EiMD04,ZHDK14}, bilayer graphene \cite{AbCh09}, or ionic plasmas \cite{M98}, other relevant perspectives deal with the inclusion of an ionic hard core, which would 
frustrate several of the arrangements put forward, the study of disordered or patterned substrates, as well as the analysis of dynamical processes and elementary excitations. 

\section*{Acknowledgment}
M.A. and G.K. gratefully acknowledge financial support by the Austrian Science Foundation (FWF) under projects Nos. P23910-N16 and F41 (SFB ViCoM) and by E-CAM, an e-infrastructure center of 
excellence for software, training and consultancy in simulation and modelling funded by the EU (Proj. No. 676531). L.\v{S}. acknowledges support from grant VEGA 2/0015/15. All authors acknowledge 
financial support from the projects PHC-Amadeus-2012 and 2015 (project numbers 26996UC and 33618YH), Projekt Amad\'ee (project numbers FR 10/2012 and FR 04/2015), and funding by Investissement 
d'Avenir LabEx PALM (grant ANR-10-LABX-0039).


\begin{thebibliography}{10}
\bibitem{rque23}
Due to symmetry considerations, we can reduce the range of $A$ to $[-1, 1]$. Excluding further the trivial case $A \in [-1, 0]$ where the charges stick to plate 1 while forming 
a two-dimensional triangular (hexagonal) lattice (the so called structure I in the phase diagram), it is sufficient to restrict our investigations to $A \in [0, 1]$.
\bibitem{rque10}
This choice (with $L=4$, see \cite{suppl}), is not the most suited for studying structure V$_x$, hence the corresponding somewhat multi-colored zone in Fig. \ref{fig:phd}. 
The parameters $\Psi_n^{(L)}$ with $L=1$ and $L=2$ would be more relevant here, but would  delineate the same region of existence. It should be kept in mind however that structures 
V$_x$ are expected at large distance $\eta$, and are not of particular interest as such.
\bibitem{rque2006}
Such structures can be defined by their symmetry group \cite{suppl}, by their large 6-fold bond orientational order parameters for both layers, their significant 4 or 8-fold order 
for layer 1, and 5 or 10-fold order for layer 2 due to the distortions of the triangular lattices.
\bibitem{rque25}
The curve that separates undercharged ($x < x_{\rm neutr}$) from overcharged ($x > x_{\rm neutr}$) states coincides with the lower boundary of structures II, III and IV in Fig. \ref{fig:phd}.
\bibitem{rque52}
For an Archimedean tiling, all vertices are equivalent and the underlying polygons are regular.
\bibitem{rque30}
Fig. \ref{fig:phd} does not clearly distinguish between \smash{S$_2$} and P patterns. Making use of parameters \smash{$\Psi_n^{(2)}$} clarifies the boundaries \cite{rque10}.
\end{thebibliography}
\end{document}